\documentclass[seceq,supplement]{ptptex}

\usepackage{graphicx}

\title{
Amplitudes in Noncritical Dimensions\\
 and Dimensional Regularization
}

\author{
Koichi \textsc{Murakami}$^{1,}$\footnote{%
  e-mail: koichimurakami71@gmail.com}
and 
Nobuyuki \textsc{Ishibashi}$^{2,}$\footnote{%
  e-mail: ishibash@het.ph.tsukuba.ac.jp}
}

\inst{
${}^{1}$Okayama Institute for Quantum Physics, 
Kyoyama 1-9-1, Kita-ku, \\
Okayama 700-0015, Japan\\
${}^{2}$Institute of Physics, University of Tsukuba,
Tsukuba, Ibaraki 305-8571, Japan
}

\abst{
We study how the dimensional regularization works in the light-cone
gauge string field theory. We show that it is not necessary to add a contact
term to the string field theory action as a counter term
in this regularization at least at the tree level.
We also investigate the one-loop amplitudes of the 
bosonic theory in noncritical dimensions
and show that they are modular invariant.

}

\begin{document}

\maketitle

\section{Introduction}
In the light-cone gauge NSR superstring field theory,
there exist divergences caused by the colliding worldsheet supercurrents
inserted at the interaction points.
In order to deal with the divergences,
we have proposed a dimensional regularization scheme
in this theory.\cite{Baba:2009kr,Baba:2009zm,Ishibashi}
\ In the conformal gauge, string theories in noncritical
dimensions which appear in the process of the dimensional regularization
correspond to the worldsheet theories with nonstandard
longitudinal part. This is an interacting CFT
that we call the $X^{\pm}$ CFT.\cite{Ishibashi,Baba:2009ns,Baba:2009fi}
\ Using this CFT, we can rewrite the tree level amplitudes
of the light-cone gauge superstring field theory
into a BRST invariant form.\cite{Baba:2009zm,Ishibashi:2010nq}
\ In the conformal gauge formulation, the vertex operators
in the Ramond sector
involve the spin fields in the $X^{\pm}$ CFT.
Since the $X^{\pm}$ CFT is an interacting theory, it is not
straightforward to construct the spin fields.
In Ref.~\citen{Ishibashi:2010nq},
we have formulated a free field description of the $X^{\pm}$ CFT
combined with the reparametrization ghosts,
and constructed the spin fields via the free variables.
Nevertheless, in the above scheme
there are no states that satisfy the level matching condition
in the (R,NS) and the (NS,R) sectors
and thus no spacetime fermions exist in the regularized
theory.\cite{Ishibashi:2010nq}
\ We need therefore modify the scheme.
A way of modification will be discussed elsewhere.

In this presentation, we would like to
study how the dimensional regularization works
in the amplitudes of the light-cone gauge superstring
field theory.
Among other things, we show that
we obtain the results of the first-quantized formulation
without adding any contact terms to the string field theory
action as counter terms.
It is an important question whether our regularization scheme
is applicable beyond the tree level amplitudes.
As a first step towards this question,
we also evaluate the one-loop amplitudes
in the light-cone gauge bosonic string field theory
in noncritical dimensions.
We show that the amplitudes are indeed modular invariant
and can be recast into a BRST invariant form.\cite{loop}

\section{Free field description  of supersymmetric $X^{\pm}$ CFT}
\label{sec:freefields}

\subsection{Free variables}

The supersymmetric $X^{\pm}$ CFT is described by using the superfield
variables
\begin{equation}
\mathcal{X}^{\pm} (\mathbf{z},\bar{\mathbf{z}})
 \equiv X^{\pm} (z,\bar{z}) + i \theta \psi^{\pm} (z)
       + i\bar{\theta} \tilde{\psi}^{\pm}(\bar{z})
       + i\theta \bar{\theta} F^{\pm} (z,\bar{z})~.
\end{equation}
The Virasoro central charge of this CFT is 
$\hat{c}=12-d$.\cite{Baba:2009fi}
\ It follows that together with the super-reparametrization ghosts
$B(\mathbf{z}) \equiv \beta (z) +\theta b(z)$,
$C(\mathbf{z}) \equiv c(z) + \theta \gamma (z)$,
and the transverse sector, the total system becomes a
superconformal field theory of vanishing
Virasoro central charge, and thus the nilpotent 
BRST charge $Q_{\mathrm{B}}$ can be constructed.

We can define the superfields $\mathcal{X}^{+}$,
$\mathcal{X}^{\prime -}$ and the ghosts
$B'$, $C'$, $\tilde{B}'$, $\tilde{C}'$,
which satisfy the OPE's for the free variables,
as\cite{Ishibashi:2010nq}
\begin{eqnarray}
&  & 
\qquad \qquad   B^{\prime}\left(\mathbf{z}\right)
      \equiv \left(D\Theta^{+}\right)^{2\alpha}
              B\left(\mathbf{z}\right)\ ,
\qquad
  C^{\prime}\left(\mathbf{z}\right)
   \equiv\left(D\Theta^{+}\right)^{-2\alpha}
          C\left(\mathbf{z}\right)\ ,
\nonumber \\
 &  & 
\mathcal{X}^{\prime-}\left(\mathbf{z},\bar{\mathbf{z}}\right)
   \equiv
     \mathcal{X}^{-}\left(\mathbf{z},\bar{\mathbf{z}}\right)
\nonumber \\
 &  & \hphantom{\mathcal{X}^{\prime-}
                \left(\mathbf{z},\bar{\mathbf{z}}\right)
                \equiv}
     {}+\alpha \left[
          \partial D \left(\Sigma^{\prime}+\frac{1}{2}\Phi\right)
          \frac{\Theta^{+}}{\left(D\Theta^{+}\right)^{3}}
          -\partial \left(\Sigma^{\prime}+\frac{1}{2}\Phi\right)
           \left(\frac{1}{\left(D\Theta^{+}\right)^{2}}
                  +\frac{\partial\Theta^{+}\Theta^{+}}
                        {\left(D\Theta^{+}\right)^{4}}
           \right)\right.
\nonumber \\
 &  & \hphantom{\mathcal{X}^{\prime-}
                \left(\mathbf{z},\bar{\mathbf{z}}\right)
                \equiv-\alpha]\quad}
   {}-D\left(\Sigma^{\prime}+\frac{1}{2}\Phi\right)
      \left(\frac{\partial\Theta^{+}}{\left(D\Theta^{+}\right)^{3}}
             +\frac{\partial D\Theta^{+}\Theta^{+}}
                    {\left(D\Theta^{+}\right)^{4}}
       \right)
\nonumber \\
 &  & \hphantom{\mathcal{X}^{\prime-}
                \left(\mathbf{z},\bar{\mathbf{z}}\right)
                \equiv-\alpha]\ }
    \left.
    {}+\mathrm{c.c.}
     \vphantom{\frac{\Theta^{+}}{\left(D\Theta^{+}\right)^{3}}}
    \right]~.
\label{eq:Xprime-}
\end{eqnarray}
Here
\begin{eqnarray}
&&
\Theta^{+} (\mathbf{z}) 
 \equiv \frac{D\mathcal{X}^{+}}
            {(\partial \mathcal{X}^{+})^{\frac{1}{2}}}
        (\mathbf{z})~,
\qquad
\Phi (\mathbf{z},\bar{\mathbf{z}}) \equiv 
  \ln \left[ -4 (D\Theta^{+})^{2} (\bar{D}\tilde{\Theta}^{+})^{2}
      \right]~,
\nonumber \\
&&
\Sigma^{\prime}\left(\mathbf{z}\right) 
\equiv 
\sigma^{\prime}\left(z\right)-\phi^{\prime}\left(z\right)
  -\theta \beta^{\prime}c^{\prime}\left(z\right)~,
\qquad
\alpha=\frac{d-10}{8}~,
\end{eqnarray}
where $\sigma^{\prime}$ and $\phi^{\prime}$ are defined so that
$\partial\sigma^{\prime}=c^{\prime}b^{\prime}$ and 
\begin{equation}
\beta^{\prime}(z)=e^{-\phi^{\prime}}\partial\xi^{\prime}(z)~,
\qquad
\gamma^{\prime}(z)=\eta^{\prime}e^{\phi^{\prime}}(z)~.
\label{eq:betaprime-bosonize}
\end{equation}
We note that 
$CB\left(\mathbf{z}\right)
  =C^{\prime}B^{\prime}\left(\mathbf{z}\right)
  =-D\Sigma^{\prime}\left(\mathbf{z}\right)$.


It is also possible to express 
$\mathcal{X}^{+}$,
$\mathcal{X}^{-}$, $B$, $C$, $\tilde{B}$, $\tilde{C}$ in terms of 
the free variables.

\subsection{Correlation functions in terms of free variables}

Now that all the fields of the theory can be expressed in terms of
the free variables, it should be possible to describe the theory
using the free variables.
We denote correlation functions on the complex plane in
the superconformal field theory consisting of $\mathcal{X}^{\pm}$
and the ghosts $B$, $C$, $\tilde{B}$, $\tilde{C}$ by
$\langle~\cdots~\rangle_{\mathcal{X}^{\pm},B,C}$, and
those of the free primed variables by
$\langle~\cdots~\rangle_{\mathrm{free}}$.
The correlation functions that we are interested in are
of the form
\begin{equation}
\left\langle 
  \left| e^{3\sigma-2\phi}(\infty) \right|^{2}
  \phi_{1}\phi_{2}\cdots\phi_{n}
  \prod_{r=1}^{N}e^{-ip_{r}^{+}\mathcal{X}^{-}}
                  \left(\mathbf{Z}_{r},\bar{\mathbf{Z}}_{r}\right)
\right\rangle _{\mathcal{X}^{\pm},B,C}~,
\label{eq:unprimedcorrelation}
\end{equation}
where the ghosts are bosonized in the usual way and 
$\left|e^{3\sigma-2\phi}\left(\infty\right)\right|^{2}$
is inserted to soak up the ghost zero modes.
We note that with the insertion of 
$\prod_{r=1}^{N}   e^{-ip^{+}_{r}\mathcal{X}^{-}}
                   (\mathbf{Z}_{r},\bar{\mathbf{Z}}_{r})$,
the superfield
$\mathcal{X}^{+}(\mathbf{z},\bar{\mathbf{z}})$ acquires
an expectation value
$-\frac{i}{2} \left( 
   \rho (\mathbf{z}) + \bar{\rho} (\bar{\mathbf{z}})
   \right)$,
where
 $\rho (\mathbf{z}) \equiv \sum_{r=1}^{N} 
   \alpha_{r} \ln (\mathbf{z}-\mathbf{Z}_{r})$
is the super Mandelstam mapping.

At a first glance, one might expect that the correlation
function~(\ref{eq:unprimedcorrelation}) should be expressed
in terms of the free variables as
\begin{eqnarray}
&&
\left\langle 
  \left| e^{3\sigma-2\phi}(\infty) \right|^{2}
  \phi_{1}\phi_{2}\cdots\phi_{n}
  \prod_{r=1}^{N}e^{-ip_{r}^{+}\mathcal{X}^{-}}
                  \left(\mathbf{Z}_{r},\bar{\mathbf{Z}}_{r}\right)
\right\rangle _{\mathcal{X}^{\pm},B,C}
\nonumber \\
&& \quad \stackrel{?}{=}
\left\langle 
  \left| e^{3\sigma-2\phi}(\infty) \right|^{2}
  \phi_{1}\phi_{2}\cdots\phi_{n}
  \prod_{r=1}^{N}e^{-ip_{r}^{+}\mathcal{X}^{-}}
                  \left(\mathbf{Z}_{r},\bar{\mathbf{Z}}_{r}\right)
\right\rangle _{\mathrm{free}}~,
\label{eq:naive}
\end{eqnarray}
on the right hand of which all the fields are considered
to be expressed in terms of the free variables
by using the relations (\ref{eq:Xprime-}).
This would hold if the relations (\ref{eq:Xprime-}) were not
singular anywhere on the complex plane.
Nevertheless, the expectation values of supercovariant derivatives
of $\mathcal{X}^{+}$
can have zeros and poles,
and thus eq.(\ref{eq:naive}) is not true as it is.
We need therefore insert operators
at $\mathbf{z}=\tilde{\mathbf{z}}_{I}$, $\mathbf{Z}_{r}$
and $\infty$ reflecting the singular behaviors of
the expectation values of supercovariant derivatives
of $\mathcal{X}^{+}$.
Here $\tilde{\mathbf{z}}_{I}$ $(I=1,\ldots, N-2)$
denotes the points determined by
$\partial \rho (\tilde{\mathbf{z}}_{I})
  = \partial D \rho (\tilde{\mathbf{z}}_{I})=0$.
We find that the resultant formula becomes\cite{Ishibashi:2010nq}
\begin{eqnarray}
 &  & 
  \left\langle 
    \left|e^{3\sigma-2\phi}\left(\infty\right)\right|^{2}
    \phi_{1}\left(\mathbf{z}_{1},\bar{\mathbf{z}}_{1}\right)
    \phi_{2}\left(\mathbf{z}_{2},\bar{\mathbf{z}}_{2}\right)
     \cdots
    \phi_{n}\left(\mathbf{z}_{n},\bar{\mathbf{z}}_{n}\right)
    \prod_{r=1}^{N}
             e^{-ip_{r}^{+}\mathcal{X}^{-}}
             \left(\mathbf{Z}_{r},\bar{\mathbf{Z}}_{r}\right)
  \right\rangle _{\mathcal{X}^{\pm},B,C}
\nonumber \\
 &  & \quad
   =\left\langle \mathcal{R}
          \phi_{1}\left(\mathbf{z}_{1}\bar{\mathbf{z}}_{1}\right)
          \phi_{2}\left(\mathbf{z}_{2},\bar{\mathbf{z}}_{2}\right)
           \cdots
          \phi_{n}\left(\mathbf{z}_{n},\bar{\mathbf{z}}_{n}\right)
          \prod_{I=1}^{N-2}\mathcal{O}_{I}
       \right.
\nonumber \\
 &  & \quad
      \hphantom{\propto\quad}
   \left. \times
    \prod_{r=1}^{N}
        \left[ \mathcal{S}_{r}  \left|\alpha_{r}\right|^{-\alpha}
               \left|e^{\alpha\Sigma^{\prime}}(\mathbf{Z}_{r})
                \right|^{2}
               e^{-ip_{r}^{+}\mathcal{X}^{\prime-}
                   +i \frac{\alpha}{2 p_{r}^{+}}\mathcal{X}^{+}}
                       \left(\mathbf{Z}_{r},\bar{\mathbf{Z}}_{r}\right)
         \right]
     \right\rangle _{\mathrm{free}},
\label{eq:supergeneral}
\end{eqnarray}
up to a numerical proportionality constant.
The operators $\mathcal{O}_{I}$, $\mathcal{R}$ and
$\mathcal{S}_{r}$ inserted on the right hand side
are defined as
\begin{subequations}
\begin{eqnarray}
\mathcal{O}_{I} 
  & \equiv & 
     \left| \oint_{\tilde{\mathbf{z}}_{I}} \frac{d\mathbf{z}}{2\pi i}
                D\Phi
       \left[
         1+\frac{\alpha}{12}
           \frac{\partial^{3}D\mathcal{X}^{+}D\mathcal{X}^{+}}
                {\left(\partial^{2}\mathcal{X}^{+}\right)^{2}}
          +\alpha \left(\frac{\alpha}{32}-\frac{1}{8}\right)
           \frac{\partial^{3}\mathcal{X}^{+}\partial^{2}
                  D\mathcal{X}^{+}D\mathcal{X}^{+}}
                {\left(\partial^{2}\mathcal{X}^{+}\right)^{3}}
\right.\right.\nonumber \\
&  & \hphantom{\oint_{\tilde{\mathbf{z}}_{I}}
                \frac{d\mathbf{z}}{2\pi i}
                D\Phi\qquad}
 {}-\frac{\alpha^{2}}{8}
          \frac{\partial^{3}\mathcal{X}^{+}D\mathcal{X}^{+}}
               {\left(\partial^{2}\mathcal{X}^{+}\right)^{2}}
          D\Sigma^{\prime}
           +\frac{\alpha^{2}}{8}
            \frac{\partial^{2}D\mathcal{X}^{+}D\mathcal{X}^{+}}
                 {\left(\partial^{2}\mathcal{X}^{+}\right)^{2}}
            \partial\Sigma^{\prime}
\nonumber \\
 &  & \hphantom{\oint_{\tilde{\mathbf{z}}_{I}}
                 \frac{d\mathbf{z}}{2\pi i}
                 D\Phi\qquad}
   \left.\left.
   {}-\frac{\alpha^{2}}{2}
      \frac{D\mathcal{X}^{+}}{\partial^{2}\mathcal{X}^{+}}
      \partial\Sigma^{\prime} D\Sigma^{\prime}
    \vphantom{\frac{\partial^{3}D\mathcal{X}^{+}D\mathcal{X}^{+}}
                   {\left(\partial^{2}\mathcal{X}^{+}\right)^{2}}
              }
    \right]
    \frac{e^{-\alpha\Sigma^{\prime}}}
         {\left(\partial^{2}\mathcal{X}^{+}\right)^{\frac{\alpha}{4}}}
     \right|^{2}~,
\\
\mathcal{R}
 &\equiv&
  \left| \oint_{\infty}\frac{d\mathbf{z}}{2\pi i}
           D\Phi
           \left(D\Theta^{+}\right)^{2\alpha}
           \left(1+\theta\gamma b\right)
           e^{3\sigma^{\prime}-2\phi^{\prime}}\left(z\right)
 \right|^{2}~,
\\
\mathcal{S}_{r}
 & \equiv& \oint_{\tilde{\mathbf{z}}_{I^{\left(r\right)}}}
            \frac{d\mathbf{z}}{2\pi i}
            D\Phi
         \oint_{\bar{\tilde{\mathbf{z}}}_{I^{\left(r\right)}}}
            \frac{d\bar{\mathbf{z}}}{2\pi i}
            \bar{D}\tilde{\Phi}
           \, e^{-i \frac{\alpha}{2 p_{r}^{+}}\mathcal{X}^{+}}
                      \left(\mathbf{z},\bar{\mathbf{z}}\right)~.
\label{eq:mathcalSr}
\end{eqnarray}
\end{subequations}
One can show that these operators are conformal invariant.

\section{Vertex operators}

\subsection{Neveu-Schwarz sector}

Let us consider the left-moving part of a state in the (NS,NS) sector
of the light-cone gauge superstrings,
\begin{equation}
\alpha_{-n_{1}}^{i_{1}} \cdots \psi^{j_{1}}_{-s_{1}} \cdots
 |\vec{p}\rangle_{L}~.
\label{eq:LCstate-NS}
\end{equation}
Here $n_{i}$ are positive integers and $s_{i}$ are positive
half odd integers.
$|\vec{p}\rangle_{L}$ is the state corresponding to
the operator $e^{i\vec{p}\cdot\vec{X}_{L}}$,
where $\vec{p}$ denotes the transverse $(d-2)$-momentum and
$\vec{X}_{L}$ denotes the left-moving part of the transverse
bosonic matter fields.
The left-moving BRST invariant vertex operator in the conformal gauge
corresponding to this state is given as\cite{Baba:2009zm}
\begin{equation}
V^{(-1)}_{L} \equiv e^{\sigma}e^{-\phi} V^{\mathrm{DDF}}_{L}~,
\label{eq:VL-1}
\end{equation}
where $V^{\mathrm{DDF}}_{L}$ denotes the DDF vertex
operator corresponding to the state (\ref{eq:LCstate-NS})
with the momentum $p^{-}$ in the $-$ direction shifted as
$p^{-} + \frac{\alpha}{2 p^{+}}$.\cite{Ishibashi:2010nq}
\ The free field expression 
$V_{L}^{\prime\left(-1\right)}$ for $V_{L}^{\left(-1\right)}$
can be obtained by using eq.(\ref{eq:supergeneral}) to be
\begin{equation}
V^{\prime (-1)}_{L} 
\equiv e^{(1+\alpha)\sigma'} e^{-(1+\alpha)\phi'}
      V^{\prime\mathrm{DDF}}_{L}~,
\label{eq:Vprime-1}
\end{equation}
where $V^{\prime\mathrm{DDF}}_{L}$
is the free field expression for $V^{\mathrm{DDF}}_{L}$.

\subsection{Ramond sector}
We turn to the left-moving part of the state in the Ramond sector
of the light-cone gauge superstring,
\begin{equation}
\alpha_{-n_{1}}^{i_{1}}\cdots\psi_{-m_{1}}^{j_{1}}\cdots
   \left|\vec{p},\vec{s}\right\rangle _{L}\ ,
\label{eq:LCstate-Ramond}
\end{equation}
where $n_{i}$ and $m_{i}$ are positive integers, and 
$\left|\vec{p},\vec{s}\right\rangle _{L}$
is the state corresponding to the operator 
$e^{i\vec{p}\cdot\vec{X}_{L}+i\vec{s}\cdot\vec{H}}$.
Here $\vec{H}=(H^{A})$ $(A=1,\ldots,\frac{d-2}{2})$ are
defined by using the transverse matter fermions as 
\begin{equation}
e^{\pm iH^{A}}
 =\frac{1}{\sqrt{2}}\left(\psi^{2A-1}\pm i\psi^{2A}\right)~,
\end{equation}
and $\vec{s}=(s^{A})$ with $s^{A}=\frac{1}{2}$ or $-\frac{1}{2}$.
In order to express the BRST invariant vertex operator for 
the state (\ref{eq:LCstate-Ramond}) in the Ramond sector,
we need use the free fields as explained in Introduction. 
Mimicking $V_{L}^{\prime\left(-1\right)}$ in eq.(\ref{eq:Vprime-1}),
we construct
\begin{eqnarray}
V_{L}^{\prime\left(-\frac{3}{2}\right)}
& \equiv & 
  e^{(1+\alpha)\sigma'}e^{-\left(\frac{3}{2}+\alpha\right) \phi'}
  e^{\frac{i}{2} H'} e^{\vec{s}_{r}\cdot \vec{H}}
  V^{\prime \mathrm{DDF}}_{L}~,
\nonumber \\
V_{L}^{\prime\prime\left(-\frac{1}{2}\right)}\left(z\right) 
& \equiv & A_{-n_{1}}^{i_{1}} \cdots B_{-m_{1}}^{j_{1}} \cdots
  e^{(1+\alpha)\sigma'}e^{-\left(\frac{1}{2}+\alpha\right) \phi'}
  e^{-\frac{i}{2} H'} e^{\vec{s}_{r}\cdot \vec{H}}
  V^{\prime \mathrm{DDF}}_{L}~,
\label{eq:VDDF-L-prime}
\end{eqnarray}
where 
$H'$ is defined by
the bosonization of the free fermions as
\begin{equation}
\psi^{+}= e^{iH'}~,\qquad
\psi^{\prime -} = - e^{-iH'}~.
\end{equation}

Let $V^{\left(-\frac{3}{2}\right)}_{L}$ be the vertex operator
in the CFT for unprimed variables which corresponds to
$V^{\prime \left(-\frac{3}{2}\right)}_{L}$
in the free field description.
Although the explicit form of 
$V^{\left(-\frac{3}{2}\right)}_{L}$
is complicated,
without it we can read off properties of
$V^{\left(-\frac{3}{2}\right)}_{L}$ from the free field form 
$V^{\prime \left(-\frac{3}{2}\right)}_{L}$.
In particular, $V^{\left(-\frac{3}{2}\right)}_{L}$
is a BRST invariant operator of 
$-\frac{3}{2}$ picture.\cite{Ishibashi:2010nq}
\ Contrastingly, $V^{\prime\prime \left(-\frac{1}{2}\right)}_{L}$
does not correspond to a BRST invariant vertex operator.
We can obtain the left-moving BRST invariant vertex operator 
$V_{L}^{\left(-\frac{1}{2}\right)}$
of $-\frac{1}{2}$ picture by applying to 
$V_{L}^{\left(-\frac{3}{2}\right)}$
the picture changing operator 
$X(z)\equiv \{Q_{\mathrm{B}}\,,\, \xi (z)\}$:
\begin{equation}
V_{L}^{\left(-\frac{1}{2}\right)}\left(0\right)
   \equiv  \lim_{z\to0} X(z)
               V_{L}^{\left(-\frac{3}{2}\right)} \left(0\right)
   = \lim_{z\to0} \left(-e^{\phi}T_{F}^{X}\right)(z)
        V_{L}^{\left(-\frac{3}{2}\right)}\left(0\right)\ .
\label{eq:VLr12}
\end{equation}
We define $V_{L}^{\prime\left(-\frac{1}{2}\right)}$ to be the free
field version of $V_{L}^{\left(-\frac{1}{2}\right)}$. 

One can obtain the BRST invariant right-moving parts 
$V_{R}^{\left(-\frac{3}{2}\right)}\left(\bar{z}\right)$,
$V_{R}^{\left(-\frac{1}{2}\right)}\left(\bar{z}\right)$
and their free field versions in the same way.


\section{Amplitudes}

The tree level $N$-string amplitudes $\mathcal{A}_{N}$ are perturbatively
computed, by starting from the string field theory action.
We obtain\cite{Ishibashi}
\begin{equation}
\mathcal{A}_{N}
 =\left(4ig\right)^{N-2}
  \int\left(\prod_{\mathcal{I}=1}^{N-3}
              \frac{d^{2}\mathcal{T}_{\mathcal{I}}}{4\pi}\right)
   F_{N}\left(\mathcal{T}_{\mathcal{I}},
              \bar{\mathcal{T}}_{\mathcal{I}}\right),
\label{eq:AN}
\end{equation}
where $\mathcal{T}_{\mathcal{I}}$ denotes the complex Schwinger
parameter of the $\mathcal{I}$th internal propagator 
$(\mathcal{I}=1,\ldots,N-3)$,
which are the $N-3$ complex moduli parameters of the amplitude
$\mathcal{A}_{N}$. As was discussed in Ref.~\citen{Baba:2009zm},
on the right hand side the integration region is taken to cover the
whole moduli space and the integrand $F_{N}$ is described by the
correlation function of the superconformal field theory for the light-cone
gauge superstrings on the $z$-plane:
\begin{eqnarray}
F_{N}\left(\mathcal{T}_{\mathcal{I}},
           \bar{\mathcal{T}}_{\mathcal{I}}\right)
 & = & \left(2\pi\right)^{2}
       \delta\left(\sum_{r=1}^{N}p_{r}^{+}\right)
       \delta\left(\sum_{r=1}^{N}p_{r}^{-}\right)
       \mathrm{sgn}\left(\prod_{r=1}^{N}\alpha_{r}\right)
       e^{-\frac{d-2}{16}\Gamma}\nonumber \\
 &  & \qquad\times
    \left\langle \prod_{I=1}^{N-2}
                   \left|\left(\partial^{2}\rho\right)^{-\frac{3}{4}}
                         T_{F}^{\mathrm{LC}}\left(z_{I}\right)
                   \right|^{2}
                \prod_{r=1}^{N}V_{r}^{\mathrm{LC}}
    \right\rangle _{X^{i}}~.
\label{eq:FN}
\end{eqnarray}
Here $e^{-\frac{d-2}{16}\Gamma}$ originates from
the conformal anomaly,\cite{Mandelstam:1985ww} \ and 
$V_{r}^{\mathrm{LC}}$
denotes the vertex operators for the $r$th external string in the
light-cone gauge. 
In the following
we will consider the case
in which $V_{r}^{\mathrm{LC}}$ $(r=1,\ldots,2f)$ are 
in the (R,R) sector
and $V_{r}^{\mathrm{LC}}$ $(r=2f+1,\ldots,N)$ are
in the (NS,NS) sector.

We would like to rewrite the light-cone gauge 
expression~(\ref{eq:AN}) of the amplitudes
into a BRST invariant form,
adding the longitudinal variables in the $X^{\pm}$ CFT
and the super-reparametrization ghosts.
A key ingredient of this rewriting 
is to express the anomaly contribution
$e^{-\frac{d-2}{16}\Gamma}$ in eq.(\ref{eq:FN})
in the form of the correlation function
of the worldsheet superconformal field theory consisting of
the $X^{\pm}$ CFT and the ghosts.
It is straightforward to show that the quantity
which appears on the right hand side of eq.(\ref{eq:FN}) can be expressed
as a correlation function in the system of the free variables defined
in \S\ref{sec:freefields}:
\begin{eqnarray}
\lefteqn{
   (2\pi)^{2} \delta\left(\sum_{r=1}^{N}p_{r}^{+}\right)
              \delta\left(\sum_{r=1}^{N}p_{r}^{-}\right)
   e^{-\frac{d-2}{16}\Gamma}
   \prod_{I=1}^{N-2}
       \left|\partial^{2}\rho\left(z_{I}\right)\right|^{-\frac{3}{2}}
   \prod_{r=1}^{N} V_{r}^{\mathrm{LC}}
}  \nonumber \\
&  & \sim
   \left\langle 
      \left|(\partial\rho)^{1+\alpha}e^{\sigma^{\prime}}(\infty)
      \right|^{2}
      \prod_{I=1}^{N-2}
         \left|\frac{e^{-\left(1+\alpha\right)
                         \left(\sigma^{\prime}-\phi^{\prime}\right)}}
                    {\left(\partial^{2}\rho\right)^{1+\frac{\alpha}{4}}}
                 (z_{I})
          \right|^{2}
       \prod_{r=1}^{f}
           \left( |\alpha_{r}|^{-\alpha}
                  V_{r}^{\prime\left(-\frac{3}{2},-\frac{3}{2}\right)}
                      \left(Z_{r},\bar{Z}_{r}\right)
            \right)
     \right.
\nonumber \\
 &  & \hphantom{\sim\quad}
   \left.\times
   \prod_{r=f+1}^{2f}
         \left(|\alpha_{r}|^{-\alpha}
               V_{r}^{\prime\prime\left(-\frac{1}{2},-\frac{1}{2}\right)}
                   \left(Z_{r},\bar{Z}_{r}\right)
          \right)
    \prod_{r=2f+1}^{N}
          \left(|\alpha_{r}|^{-\alpha}
                V_{r}^{\prime\left(-1,-1\right)}
                    \left(Z_{r},\bar{Z}_{r}\right)
          \right)
  \right\rangle _{\mathrm{free}}.
\nonumber\\
\label{eq:VDDF-VLC-RNS}
\end{eqnarray}
Here 
\begin{eqnarray}
V_{r}^{\prime\left(-1,-1\right)}(Z_{r},\bar{Z}_{r})
 & \equiv & V_{L,r}^{\prime\left(-1\right)}(Z_{r})
            V_{R,r}^{\prime\left(-1\right)}(\bar{Z}_{r})~,
\nonumber \\
V_{r}^{\prime\left(-\frac{3}{2},-\frac{3}{2}\right)}(Z_{r},\bar{Z}_{r})
  & \equiv & |\alpha_{r}|^{-(\alpha+1)}
             V_{L,r}^{\prime\left(-\frac{3}{2}\right)}(Z_{r})
             V_{R,r}^{\prime\left(-\frac{3}{2}\right)}(\bar{Z}_{r})~,
\nonumber \\
V_{r}^{\prime\prime\left(-\frac{1}{2},-\frac{1}{2}\right)}
    (Z_{r},\bar{Z}_{r}) 
  & \equiv & |\alpha_{r}|^{\alpha+1}
             V_{L,r}^{\prime\prime\left(-\frac{1}{2}\right)}(Z_{r})
             V_{R,r}^{\prime\prime\left(-\frac{1}{2}\right)}
                    (\bar{Z}_{r})~,
\end{eqnarray}
where $V_{L,r}^{\prime\left(-1\right)}$, 
$V_{L,r}^{\prime\left(-\frac{3}{2}\right)}$
and $V_{L,r}^{\prime\prime\left(-\frac{1}{2}\right)}$ are the vertex
operators $V_{L}^{\prime\left(-1\right)}$, 
$V_{L}^{\prime\left(-\frac{3}{2}\right)}$
and $V_{L}^{\prime\prime\left(-\frac{1}{2}\right)}$ 
defined in eqs.(\ref{eq:Vprime-1})
and (\ref{eq:VDDF-L-prime}) for the $r$-th external string, and
similarly for the right moving sector ones 
$V_{R,r}^{\prime\left(-1\right)}$,
$V_{R,r}^{\prime\left(-\frac{3}{2}\right)}$ and 
$V_{R,r}^{\prime\prime\left(-\frac{1}{2}\right)}$.

On the right hand side of eq.(\ref{eq:VDDF-VLC-RNS}), $X^{\prime-}$
appears only in the form of the vertex operator 
$e^{-ip^{+}X^{\prime-}}$,
and $\psi^{\prime-},\tilde{\psi}^{\prime-}$ do not appear. 
Therefore
we can replace $X^{+},\psi^{+},\tilde{\psi}^{+}$ by their expectation
values $-\frac{i}{2}\left(\rho+\bar{\rho}\right),0,0$ in the correlation
function, and vice versa.
In this situation, 
without changing the value of the correlation function,
we can replace 
$V^{\prime\prime \left(-\frac{1}{2},-\frac{1}{2}\right)}_{r}$
with
$V^{\prime \left(-\frac{1}{2},-\frac{1}{2}\right)}_{r}$
that is defined
by applying the picture changing
operator $X$ and its right-moving counter part $\tilde{X}$
to $V_{r}^{\prime \left(-\frac{3}{2},-\frac{3}{2}\right)}$
as\cite{Ishibashi:2010nq}
\begin{equation}
V_{r}^{\prime \left(-\frac{1}{2},-\frac{1}{2}\right)}
 \equiv X \tilde{X} 
        V_{r}^{\prime \left(-\frac{3}{2},-\frac{3}{2}\right)}~.
\end{equation}
Besides this replacement,
we can further recast the right hand side
of eq.(\ref{eq:VDDF-VLC-RNS}) into
\begin{eqnarray}
&  & 
  \left\langle 
          \left| \left(\sum_{r}\alpha_{r}Z_{r}\right)
                 \lim_{z\to\infty}
                 e^{-2\left(\sigma^{\prime}-\phi^{\prime}\right)}
                     \left(z\right)
          \right|^{2}
          \mathcal{R}
       \right.
\nonumber \\
 &  &\quad
   \times\prod_{I=1}^{N-2}
      \left| \oint_{z_{I}} \frac{dz}{2\pi i}
               \frac{e^{-\sigma^{\prime}}}
                    {\left(\partial\rho\right)^{1+\alpha}}\left(z\right)
               \lim_{w\to z_{I}}
                   \left(\left(\partial\rho\right)^{\alpha} 
                   e^{\phi^{\prime}}\right)\left(w\right)
      \right|   \mathcal{O}_{I}
\nonumber \\
 &  & \quad
   \left.\times
    \prod_{r=1}^{N}
      \left( |\alpha_{r}|^{-\alpha}
             V_{r}^{\prime\left(p_{L,r},p_{R,r}\right)}
                    \left(Z_{r},\bar{Z}_{r}\right)
      \right)
   \right\rangle _{\mathrm{free}}~,
\label{eq:VDDF-VLC-RNS2}
\end{eqnarray}
where 
$p_{L,r},p_{R,r}=-\frac{1}{2},-1,-\frac{3}{2}$ indicate
the picture of the vertex operators.
The choice of the picture in eq.(\ref{eq:VDDF-VLC-RNS2})
is obvious from eq.(\ref{eq:VDDF-VLC-RNS}).
To eq.(\ref{eq:VDDF-VLC-RNS2})  we can easily apply
the formula~(\ref{eq:supergeneral}), and express it
using the $X^{\pm}$ CFT and the unprimed ghost fields. Substituting
it into eq.(\ref{eq:FN}), we obtain 
\begin{equation}
F_{N} 
 \sim  
  \left\langle 
      \left|\partial\rho c\left(\infty\right)\right|^{2}
      \prod_{I}\left|\oint_{z_{I}}\frac{dz}{2\pi i}
                        \frac{b}{\partial\rho}\left(z\right)
                     e^{\phi}T_{F}^{\mathrm{LC}}\left(z_{I}\right)
                \right|^{2}
      \prod_{r=1}^{N}\mathcal{S}_{r}^{-1}
      \prod_{r=1}^{N}V_{r}^{\left(p_{L,r},p_{R,r}\right)}
          \left(Z_{r},\bar{Z}_{r}\right)\right\rangle.
\label{eq:FNunprime}
\end{equation}
Here $\left\langle \cdots\right\rangle $ denotes the correlation
function of the CFT for the longitudinal 
and transverse variables and the super-reparametrization ghosts. 
$V_{r}^{\left(p_{L,r},p_{R,r}\right)}$ is
the unprimed field version of 
$V_{r}^{\prime\left(p_{L,r},p_{R,r}\right)}$
and it is BRST invariant. 
$\mathcal{S}_{r}^{-1}$ 
is defined as 
\begin{equation}
\mathcal{S}_{r}^{-1}
  \equiv
  \oint_{z_{I^{\left(r\right)}}} \frac{d\mathbf{z}}{2\pi i}
         D\Phi\left(\mathbf{z}\right)
  \oint_{\bar{z}_{I^{\left(r\right)}}} \frac{d\bar{\mathbf{z}}}{2\pi i}
         \bar{D}\Phi\left(\bar{\mathbf{z}}\right)
  e^{i \frac{\alpha}{2p_{r}^{+}}\mathcal{X}^{+}}
         \left(\mathbf{z},\bar{\mathbf{z}}\right)\ ,
\label{eq:Sinverse}
\end{equation}
which can be shown to be the inverse of $\mathcal{S}_{r}$ 
in eq.(\ref{eq:mathcalSr})
by replacing $\mathcal{X}^{+}$ by its expectation value. 

In eq.(\ref{eq:FNunprime}), the right hand side is expressed by the
variables in the conformal gauge, but it is not manifestly BRST invariant.
In order to get a BRST invariant form of the amplitudes, we would
like to show that  $e^{\phi}T_{F} (z_{I})$ 
in eq.(\ref{eq:FNunprime})
can be turned into the picture changing operator $X(z_{I})$
and thus
\begin{equation}
F_{N}  \sim 
   \left\langle \left|\partial\rho c (\infty)\right|^{2}
       \prod_{I}\left|\oint_{z_{I}}\frac{dz}{2\pi i}
                          \frac{b}{\partial\rho} (z)
                       X (z_{I})
                 \right|^{2}
        \prod_{r=1}^{N}\mathcal{S}_{r}^{-1}
        \prod_{r=1}^{N}V_{r}^{\left(p_{L,r},p_{R,r}\right)}
   \right\rangle ~.
\label{eq:FNpic}
\end{equation}
Let us introduce a nilpotent fermionic charge $Q$ defined as 
\begin{equation}
Q \equiv
  \oint\frac{dz}{2\pi i}
    \left[ -\frac{1}{4} \frac{b}{\partial\rho}
            \left(iX_{L}^{+} - \frac{1}{2} \rho \right)
           +\frac{1}{2}
            \frac{e^{-\phi}\partial\xi}{\partial\rho}\psi^{+}
     \right]  (z)~.
\end{equation}
One can show 
\begin{eqnarray}
\oint_{z_{I}}\frac{dz}{2\pi i}
  \frac{b}{\partial\rho} (z) X(z_{I}) 
& = & -\oint_{z_{I}} \frac{dz}{2\pi i}
   \frac{b}{\partial\rho} (z)
   e^{\phi}T_{F}^{\mathrm{LC}} (z_{I})
\nonumber \\
 &  & {}+\left[Q\,,\,
               \oint_{z_{I},w} \frac{dz}{2\pi i}
                 \frac{b}{\partial\rho} (z)
               \oint_{z_{I}}\frac{dw}{2\pi i}
                  \frac{A(w)}{w-z_{I}}e^{\phi} (z_{I})
          \right]
\nonumber \\
 &  & {}+\frac{1}{4}
         \oint_{z_{I},w}\frac{dz}{2\pi i}
           \frac{b}{\partial\rho} (z)
         \oint_{z_{I}}\frac{dw}{2\pi i}
           \frac{\partial\rho\psi^{-} (w)}{w-z_{I}}
           e^{\phi} (z_{I}),
\label{eq:XQexact}
\end{eqnarray}
where 
the explicit form of $A(w)$ is given in eq.(5.21)
of Ref.~\citen{Ishibashi:2010nq}.
Substituting eq.(\ref{eq:XQexact}) into the right hand side 
of eq.(\ref{eq:FNpic})
and comparing it with that of eq.(\ref{eq:FNunprime}), 
one can see that in order to prove eq.(\ref{eq:FNpic}),
one should show that the second and the third terms
on the right hand side of eq.(\ref{eq:XQexact}) do not contribute
to the correlation function.
One can prove the third term does not contribute to the correlation 
function.\cite{Baba:2009zm,Ishibashi:2010nq}
\ The second term is $Q$-exact. We can therefore prove that this is
also irrelevant, by showing that $Q$ (anti)commutes with all the
operators in the correlation function (\ref{eq:FNpic}). 
Thus we obtain the expression (\ref{eq:FNpic}) for $F_{N}$.

By deforming the contour of 
$\oint_{z_{I}}\frac{dz}{2\pi i}\frac{b}{\partial\rho} (z)$
in eq.(\ref{eq:FNpic})
as was done in Ref.~\citen{Baba:2009zm}, we can obtain a manifestly
BRST invariant form of the amplitude $\mathcal{A}_{N}$:
\begin{eqnarray}
&&
\mathcal{A}_{N} 
 \sim 
  \int \prod_{\mathcal{I}=1}^{N-3} d^{2}\mathcal{T}_{\mathcal{I}}
  \left\langle 
     \prod_{\mathcal{I}=1}^{N-3}
        \left[ \oint_{C_{\mathcal{I}}}\frac{dz}{2\pi i}
                  \frac{b}{\partial\rho}(z)
               \oint_{C_{\mathcal{I}}}\frac{d\bar{z}}{2\pi i}
                   \frac{\tilde{b}}{\bar{\partial}\bar{\rho}}(\bar{z})
          \right]
     \prod_{I} \left| X \left(z_{I}\right) \right|^{2}
\right. \nonumber \\
&& \hspace{9em} \left. \times
     \prod_{r=1}^{N} \mathcal{S}_{r}^{-1}
     \prod_{r=1}^{N}V_{r}^{\left(p_{L,r},p_{R,r}\right)}
  \right\rangle \ .
\end{eqnarray}
Here $C_{\mathcal{I}}$ denotes a contour which goes around the 
$\mathcal{I}$th internal propagator. 
 Compared with the form of the tree amplitudes in the critical case, 
the difference is in the insertions of $\mathcal{S}_{r}^{-1}$. These
insertions are peculiar to the noncritical strings.~\cite{Baba:2009ns}
\ It follows that in the limit 
$d \rightarrow 10$
the amplitudes $\mathcal{A}_{N}$ smoothly coincide with
the results of the first-quantized formalism in the critical
dimensions.
This implies that 
in our dimensional regularization scheme
we need not add any contact term interactions
to the string field theory action as counter terms.

\section{One-loop amplitudes of light-cone gauge bosonic string field
theory in noncritical dimensions}

\begin{figure}
\centerline{\includegraphics[width=11.5cm]{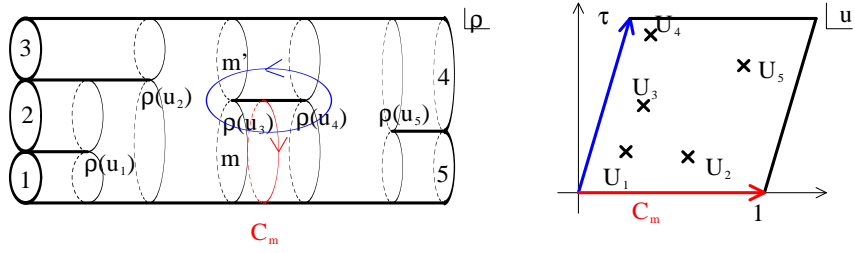}}
\caption{One-loop string diagram and the parallelogram on the $u$-plane}
\label{fig:1}
\end{figure}

We introduce the complex coordinate $\rho$ on the 
$N$-string one-loop diagram in the usual way.
We can map the $\rho$-plane to the periodic parallelogram
on the complex $u$-plane with period $1$ and $\tau$
with $\mathop{\mathrm{Im}} \tau >0$ (Fig.~\ref{fig:1}),
using the Mandelstam mapping\cite{Mandelstam:1985ww}
\begin{equation}
\rho (u) = \sum_{r=1}^{N} \alpha_{r}
 \left[ \ln \vartheta_{1} (u-U_{r}|\tau)
        - 2\pi i \frac{\mathop{\mathrm{Im}} U_{r}}
                      {\mathop{\mathrm{Im}} \tau} u \right]~.
\label{eq:Mandelstamloop}
\end{equation}
$U_{r}$ is the position of the puncture on the $u$-plane
corresponding to the $r$th external string.
We denote the $N$ interaction points
by $u_{I}$ $(I=1,\ldots, N)$, which are determined
by $\partial \rho (u_{I}) =0$.

In a similar way to the tree level case,
the one-loop amplitudes for $N$ strings
are described as correlation functions of
the worldsheet CFT on the $u$-plane.
This takes the form
\begin{equation}
\mathcal{A}_{N} \sim
 (2\pi)^{2}
      \delta^{2} \left( \sum_{r=1}^{N} p^{\pm}_{r} \right) 
 \int d^{2}\mathcal{T} \int \prod_{\mathcal{I}=1}^{N-2}
  d^{2}\mathcal{T}_{\mathcal{I}} 
  \int\frac{\alpha_{\mathrm{m}}d\alpha_{\mathrm{m}}}
           {4\pi}
      \frac{d\theta_{\mathrm{m}}}{2\pi}
      \left\langle \prod_{r=1}^{N} V_{r}^{\mathrm{LC}}
      \right\rangle
     e^{-\frac{d-2}{24} \Gamma_{\mathrm{loop}}}~.
\label{eq:ANloop}
\end{equation}
Here $\mathcal{T}$ denotes the complex Schwinger parameter
for the internal propagator of the loop part, and
$\mathcal{T}_{\mathcal{I}}$ $(\mathcal{I}=1,\ldots,N-2)$
are those for the other internal propagators.
$\alpha_{\mathrm{m}}$ and $\theta_{\mathrm{m}}$
denote the string length and the twist angle
of the internal propagator of the loop part
(Fig.~\ref{fig:1}).
$e^{-\frac{d-2}{24}\Gamma_{\mathrm{loop}}}$
denotes the contribution from the conformal 
anomaly,\cite{Mandelstam:1985ww} \ where
\begin{eqnarray}
e^{-\Gamma_{\mathrm{loop}}}
 &\equiv& \prod_{r=1}^{N} \left[ \alpha_{r}^{-2}
      e^{-2 \mathop{\mathrm{Re}} \bar{N}^{rr}_{00}} \right]
   \prod_{I=1}^{N} \left| \partial^{2} \rho (u_{I}) \right|^{-1}~,
\nonumber \\
\bar{N}^{rr}_{00} 
 &\equiv& \frac{1}{\alpha_{r}} \left[ 
       -\sum_{s\neq r} \alpha_{s} \ln \vartheta_{1} (U_{r}-U_{s}|\tau)
       + i2\pi \frac{U_{r}}{\mathop{\mathrm{Im}} \tau}
          \sum_{s=1}^{N} \alpha_{s} \mathop{\mathrm{Im}} U_{s}
     +\rho (u_{I^{(r)}}) \right]
\nonumber \\
 && {} - \ln \vartheta'_{1} (0|\tau)~. 
\end{eqnarray}

Let us study the behavior of
the amplitude $\mathcal{A}_{N}$ in eq.(\ref{eq:ANloop})
under the modular transformations
\begin{equation}
u \mapsto \frac{u}{c\tau +d}~,
\qquad
\tau \mapsto \frac{a\tau + b}{c\tau + d}~,
\label{eq:modulartrf}
\end{equation}
with $a,b,c,d \in \mathbb{Z}$
and $ad-bc=1$.
One can easily find that under
the transformation~(\ref{eq:modulartrf})
the Mandelstam mapping~(\ref{eq:Mandelstamloop})
transforms as
\begin{equation}
\rho (u) \mapsto 
\rho (u) + i\pi \frac{c}{c\tau +d}
           \sum_{r=1}^{N} \alpha_{r} U_{r}^{2}~.
\end{equation}
This leads to the modular invariance of
 $\mathcal{A}_{N}$
in eq.(\ref{eq:ANloop}).

Similarly to the tree level amplitudes,
we can describe $e^{-\frac{d-2}{24}\Gamma_{\mathrm{loop}}}$
in terms of the correlation function of
the worldsheet CFT consisting of the $X^{\pm}$ CFT
and the $bc$-ghost system.
Eventually, we can rewrite the amplitudes
$\mathcal{A}_{N}$ in eq.(\ref{eq:ANloop})
into a BRST invariant form as
\begin{eqnarray}
\mathcal{A}_{N} 
&\sim& \int d^{2}\mathcal{T} 
    \int \prod_{\mathcal{I}=1}^{N-2} d^{2}\mathcal{T}_{\mathcal{I}}
    \int \frac{\alpha_{\mathrm{m}} d\alpha_{\mathrm{m}}}{4\pi}
          \frac{d\theta_{\mathrm{m}}}{2\pi} 
  \mathop{\mathrm{Im}}\tau
\nonumber \\
&& \quad \times
\left\langle 
 \prod_{r=1}^{N} \left[ c\tilde{c}:V_{r}^{\mathrm{DDF}} 
                 e^{-\frac{d-26}{24}\frac{i}{p^{+}_{r}} X^{+}}
                    (U_{r},\bar{U}_{r}): \right]
 \prod_{r=1}^{N}  \mathcal{S}^{-1}_{(\mathrm{bosonic}),r}
\right. 
\nonumber \\
&& \qquad \quad \left. \times
 \left| \oint_{C_{\mathrm{m}}} \frac{du}{2\pi i} \frac{b}{\partial \rho}(u)
        \oint_{C_{\mathrm{m'}}} \frac{du}{2\pi i} \frac{b}{\partial \rho}(u)
 \right|^{2}
 \prod_{\mathcal{I=1}}^{N-2} \left| \oint_{C_{\mathcal{I}}}
       \frac{du}{2\pi i} \frac{b}{\partial \rho} (u)
           \right|^{2}
\right\rangle,
\end{eqnarray}
where $\mathcal{S}_{(\mathrm{bosonic}),r}^{-1}$
denotes the bosonic part of
$\mathcal{S}_{r}^{-1}$ in eq.(\ref{eq:Sinverse}).

\section{Summary and discussions}

We have seen that our dimensional regularization
scheme works well in the light-cone gauge superstring
field theory at least at the tree level.
In particular, we have shown that in our scheme
we can obtain the results of the first-quantized
formulation without introducing any contact term interactions
as counter terms.

As a first step towards the application of the dimensional
regularization to the loop level,
we have studied the one-loop amplitudes
of the light-cone gauge bosonic string field theory
in noncritical dimensions. We have shown
that the amplitudes are modular invariant
and 
can be recast into a BRST invariant form 
by using the $X^{\pm}$ CFT.
Another thing to be investigated is the Green-Schwarz formalism.
As was pointed out in Ref.~\citen{Kazama:2010ys},
our results seems to be useful in constructing the vertex operators
in the semi-light-cone gauge Green-Schwarz string theory.
In particular, the similarity transformation given 
in Ref.~\citen{Kazama:2010ys} looks similar to
the bosonic part of eq.(\ref{eq:Xprime-}).
It will be interesting to apply our results to this formulation.

\section*{Acknowledgements}
We would like to thank the organizers of the conference
for a wonderful and inspiring meeting.
This work was supported in part by
Grant-in-Aid for Scientific Research~(C) (20540247) from
the Ministry of Education, Culture, Sports, Science and
Technology (MEXT) of Japan.

%


\begin{thebibliography}{99}

\bibitem{Baba:2009kr}
  Y.~Baba, N.~Ishibashi and K.~Murakami,
  \JHEP{10,2009,035}
  [arXiv:0906.3577 [hep-th]].


\bibitem{Baba:2009zm}
  Y.~Baba, N.~Ishibashi and K.~Murakami,
  \JHEP{08,2010,102}
  [arXiv:0912.4811 [hep-th]].


\bibitem{Ishibashi}
N.~Ishibashi and K.~Murakami,
``Light-cone Gauge String Field Theory
  and Dimensional Regularization," in this volume.

\bibitem{Baba:2009ns}
  Y.~Baba, N.~Ishibashi and K.~Murakami,
  \JHEP{12,2009,010}
  [arXiv:0909.4675 [hep-th]].




\bibitem{Baba:2009fi}
  Y.~Baba, N.~Ishibashi and K.~Murakami,
  \JHEP{01,2010,119}
  [arXiv:0911.3704 [hep-th]].


\bibitem{Ishibashi:2010nq}
  N.~Ishibashi and K.~Murakami, \JHEP{01,2011,008},
  [arXiv:1011.0112 [hep-th]].

\bibitem{loop} N.~Ishibashi and K.~Murakami, work in progress.

\bibitem{Mandelstam:1985ww}
  S.~Mandelstam,
  ``The Interacting String Picture And Functional Integration,''
  Lectures given at Workshop on Unified String Theories, 
  Santa Barbara, CA, Jul 29 - Aug 16, 1985. 


\bibitem{Kazama:2010ys}
  Y.~Kazama and N.~Yokoi,
  arXiv:1008.4655 [hep-th].


\end{thebibliography}
\end{document}